\documentclass[prb,twocolumn, showpaces,preprintnumbers,amsmath,amssymb,superscriptaddress]{revtex4-1}
\usepackage{graphicx}
\usepackage{dcolumn}
\usepackage{bm}
\usepackage{color}
\usepackage{amssymb}
\usepackage{amsmath}
\usepackage{epstopdf}
\usepackage{bbding}
\usepackage[mathlines]{lineno}
\usepackage[dvipdfm, pdfstartview=FitH, CJKbookmarks=true, bookmarksnumbered=true, bookmarksopen=true, colorlinks, pdfborder=001, linkcolor=black, anchorcolor=blue, citecolor=blue]{hyperref}
\usepackage{lastpage}
\usepackage{fancyhdr}
\begin{document}
%\linenumbers
\preprint{APS/123-QED}
\title{Magnetocrystalline anisotropic effect in GdCo$_{1-x}$Fe$_x$AsO ($x = 0, 0.05$)}
\author{T. Shang}
\thanks{Present Address: Key Laboratory of Magnetic Materials and Devices $\&$ Zhejiang Province Key Laboratory of Magnetic Materials and Application
Technology, Ningbo Institute of Materials Technology and Engineering, Chinese Academy of Sciences, Ningbo 315201, China}
\affiliation{Center for Correlated Matter and Department of Physics, Zhejiang University, Hangzhou 310058, China}
\author{Y. H. Chen}
\affiliation{Center for Correlated Matter and Department of Physics, Zhejiang University, Hangzhou 310058, China}
\author{F. Ronning}
\affiliation{Los Alamos National Laboratory, Los Alamos, New Mexico 87545, USA}
\author{N. Cornell}
\affiliation{UTD-NanoTech Institute, The University of Texas at Dallas, Richardson, Texas 75083$\textendash$0688, USA}
\author{J. D. Thompson}
\affiliation{Los Alamos National Laboratory, Los Alamos, New Mexico 87545, USA}
\author{A. Zakhidov}
\affiliation{UTD-NanoTech Institute, The University of Texas at Dallas, Richardson, Texas 75083$\textendash$0688, USA}
\author{M. B. Salamon}
\affiliation{UTD-NanoTech Institute, The University of Texas at Dallas, Richardson, Texas 75083$\textendash$0688, USA}
\author{H. Q. Yuan}
\email{hqyuan@zju.edu.cn}
\affiliation{Center for Correlated Matter and Department of Physics, Zhejiang University, Hangzhou 310058, China}
\affiliation{Collaborative Innovation Center of Advanced Microstructures, Nanjing 210093, China}
\date{\today}

\begin{abstract}
From a systematic study of the electrical resistivity $\rho(T,H)$, magnetic susceptibility $\chi(T,H)$, isothermal magnetization $M(H)$ and the specific heat $C(T,H)$, a temperature-magnetic field ($T$-$H$) phase diagram has been established for GdCo$_{1-x}$Fe$_x$AsO ($x = 0$ and $0.05$) polycrystalline compounds. GdCoAsO undergoes two long-range magnetic transitions: ferromagnetic (FM) transition of Co $3d$ electrons ($T_\textup{C}^\textup{Co}$) and antiferromagnetic (AFM) transition of Gd $4f$ electrons ($T_\textup{N}^\textup{Gd}$). For the Fe-doped sample ($x=0.05$), an extra  magnetic reorientation transition takes place below $T_\textup{N}^\textup{Gd}$, which is likely associated with Co moments. The two magnetic species of Gd and Co are coupled antiferromagnetically to give rise to ferrimagnetic (FIM) behavior in the magnetic susceptibility. Upon decreasing the temperature ($T < T_\textup{C}^\textup{Co}$), the magnetocrystalline anisotropy breaks up the FM order of Co by aligning the moments with the local easy axes of the various grains, leading to a spin reorientation transition at $T_\textup{R}^\textup{Co}$. By applying a magnetic field, $T_\textup{R}^\textup{Co}$ monotonically decreases to lower temperatures, while the $T_\textup{N}^\textup{Gd}$ is relatively robust against the external field. On the other hand, the applied magnetic field pulls the magnetization of grains from the local easy direction to the field direction via a first-order reorientation transition, with the transition field ($H_\textup{M}$) increasing upon cooling the temperature.
\begin{description}
\item[PACS number(s)]
74.70.Xa,75.30.Kz,75.30.Gw
\end{description}
\end{abstract}

\maketitle
\section{\label{sec:level1}INTRODUCTION}

The interplay of $3d$ and $4f$ electrons in the FeAs-based compounds gives rise to rich physical properties, e.g., quantum phase transition, heavy fermion behavior, reentrance of superconductivity and complex magnetism.\cite{E.M.Bruning2008,Kitagawa2012,Eu122,tianCeFeAsOF} In the ZrCuSiAs-type $RE$Fe$Pn$O compounds ($RE$ = rare earth, $Pn$ = pnictogen), the 4$f$-electrons usually form an antiferromagnetic (AFM) order at very low temperatures, while the occurrence of superconductivity is mainly associated with the Fe-$3d$ electrons.\cite{kamihara2008iron, cxh2008, wang2008thorium, Ren2008, chen2008superconductivity} For instance, CeFeAsO sequentially undergoes two AFM-type transitions upon cooling from room temperature, one associated with Fe ($T_\textup{N}^\textup{Fe} \approx$ 150 K) and the other one attributed to Ce ($T_\textup{N}^\textup{Ce} \approx$ 3.4 K).\cite{chen2008superconductivity, zhao2008structural} Elemental substitutions in CeFeAsO, e.g., Fe/Co or As/P, may induce superconductivity while suppressing the magnetic order of Fe-$3d$ electrons. \cite{luo2010phase, de2010lattice, Jesche, tianCe,zhao2010effects} On the other hand, CeFePO is a paramagnetic (PM) heavy fermion metal.~\cite{E.M.Bruning2008} Evidence for a magnetic quantum phase transition was shown in CeFeAs$_{1-x}$P$_x$O, where the AFM state is separated from the PM heavy fermion state, and a ferromagnetic order develops in the intermediate doping region.~\cite{luo2010phase, de2010lattice, Jesche} These indicate that the 1111-type iron pnictides may provide a representative system to study the interplay of $4f$ and $3d$ electrons, and their emergent properties. A systematic study of the interplay between $4f$ and $3d$ electrons would help elucidate the nature of superconductivity and magnetism in iron pnictides.

In the Ce(Fe,Co)AsO series, superconductivity appears while the magnetic order at $T_\textup{N}^\textup{Fe}$ is quickly suppressed upon substituting Fe with Co.\cite{tianCe,zhao2010effects} The Ce N\'{e}el temperature slightly increases near the Fe-$3d$ magnetic instability, but shows a nearly unchanged value on further increasing the Co concentration,\cite{tianCe} indicating a strong coupling between Ce $4f$ electrons and Fe $3d$ electrons in CeFeAsO, as previously reported by neutron scattering and muon spin relaxation measurements.\cite{chi2008crystalline, maeter2009interplay} On the Co-rich side, the ferromagnetically ordered Co ions have a strong polarization effect on the AFM order of Ce moments.\cite{tianCe, sarkar2010interplay} On the other hand, in the Gd(Fe,Co)AsO series, which correspond to the case of a large $4f$-magnetic moment, Fe/Co substitution also leads to a superconducting dome near the $3d$-electron magnetic instability.\cite{tianCe} On the Co-rich side, in contrast to the Ce-case, the large moment of Gd ions is robust against the FM order of Co moments, and maintains its AFM order with a nearly unchanged N\'{e}el temperature over the entire doping range.\cite{tianCe} However, the AFM coupling between the Gd- and Co-blocks leads to ferrimagnetic behavior at low fields and a possible magnetic reorientation below $T_\textup{N}^\textup{Gd}$.\cite{tianCe,JunmuSR2011} For the case of $RE$ = Sm and Nd, both show behavior similar to the Ce compounds on the Fe-rich side.\cite{tianCe, marcinkova2010superconductivity,wang2009effects} However, on the Co-end, namely, SmCoAsO and NdCoAsO, complex magnetic properties were observed, in which the Co moments sequentially undergo an FM and an AFM transition, followed by an AFM transition of $4f$ electrons at $T_\textup{N}^\textup{Sm}$ = 5 K and $T_\textup{N}^\textup{Nd}$ = 3.5 K.\cite{awana2010magnetic,mcguire2010magnetic} By applying a magnetic field, such an FM-AFM transition quickly shifts to lower temperatures, showing a large magnetoresistance effect up to the Curie temperature.\cite{Ohta2009} However, in GdCoAsO, the magnetic susceptibility exhibits a broad maximum at temperatures below $T_\textup{C}^\textup{Co}$, being different from the FM-AFM transition in the Sm- and Nd-case, and its origin remains unclear.\cite{tianCe,JunmuSR2011,Ohta2009} Since the transitions of the Co $3d$ electrons are susceptible to the application of magnetic field in these compounds, e.g., SmCoAsO and NdCoAsO,\cite{Ohta2009} it is also interesting to tune the $d$-$f$ magnetic coupling in GdCoAsO and investigate the associated phenomena.

In this paper, we present the resultant $T$-$H$ phase diagram for GdCo$_{1-x}$Fe$_x$AsO ($x$ = 0 and 0.05) polycrystalline compounds based on systematic measurements of the physical properties. We have studied the properties of $x$ = 0, 0.05 and 0.1 (nominal values) under various magnetic fields and all these samples demonstrate similar behaviors even though the Curie temperature $T_\textup{C}^\textup{Co}$ systematically decreases with increasing $x$. For clarity, only the results of $x=0$ and 0.05 are going to be presented in the paper. For these compounds, the AFM order of Gd $4f$ electrons is relatively robust against magnetic field, while the magnetic transitions of Co blocks are sensitive to the magnetic field. The structure of this paper is organized as follows: to begin with, we will provide a brief introduction in Sec.I. Experimental methods are described in Sec.II. In Sec. III, we present the experimental results of the electrical resistivity, magnetic properties, and specific heat under various magnetic fields. Finally, Sec.IV summarizes the results of this investigation.

\section{EXPERIMENTAL DETAILS}

Polycrystalline samples of GdCo$_{1-x}$Fe$_x$AsO ($x$ = 0.0 and 0.05) were synthesized by solid-state reaction as described elsewhere.\cite{tianCe} The crystal structure of these compounds was characterized by powder x-ray diffraction (XRD) at room temperature using the PANalytical X'Pert MRD diffractometer and a graphite monochromator. The chemical compositions of the compounds were estimated by using an energy dispersive X-ray spectrometer (EDXS), from which an actual Fe concentration of $x$=0.07 and 0.12 were determined for the nominal $x$=0.05 and 0.1, respectively. We note that the nominal values of $x=0$ and 0.05 are used in this paper. Measurements of magnetic properties and specific heat were performed in a Quantum-Design magnetic property measurement system (MPMS-7T) and a physical property measurement system (PPMS-9T), respectively. Temperature dependence of the electrical resistivity was measured by a standard four-point method for temperatures $T$ = 2-300 K by using a LR700 resistance bridge combined with the PPMS temperature control system.

\section{RESULTS AND DISCUSSION}

\begin{figure}[tbp]
     \begin{center}
     \includegraphics[width=3.35in,keepaspectratio]{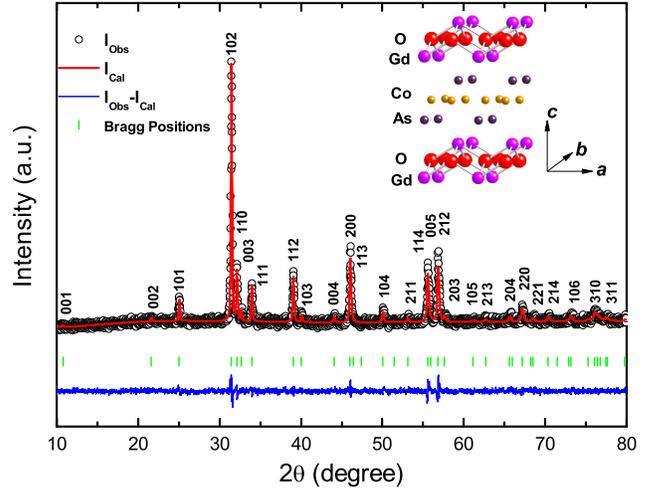}
     \end{center}
     \caption{(Color online) Room-temperature powder XRD patterns for GdCoAsO (black open circles). The solid red line is the Rietveld refinement profile, while the blue one represents the differences between the calculated and the experimental patterns. The vertical green bars mark the calculated positions of the Bragg peaks. The crystal structure of GdCoAsO is shown in the inset.}
     \label{fig1}
\end{figure}

Figure 1 presents the XRD patterns of GdCoAsO at room temperature, with GdCo$_{0.95}$Fe$_{0.05}$AsO showing similar results. The XRD patterns were analyzed by Rietveld refinement using the GSAS+EXPGUI program.\cite{Rietveld,gsas} Both samples exhibit a single phase and crystallize in the tetragonal ZrCuSiAs-type structure with space group P4/nmm, as shown in the inset of Fig. 1, which can be viewed as alternating GdO- and Co(Fe)As-layers stacking along the $c$-axis. According to the Rietveld refinement, the derived lattice parameters are $a = b = 3.937\textup{\r{A}}$, $c = 8.204\textup{\r{A}}$ for GdCoAsO, being consistent with previous results.\cite{tianCe,Ohta2009}

\begin{figure}[tbp]
     \begin{center}
     \includegraphics[width=3.3in,keepaspectratio]{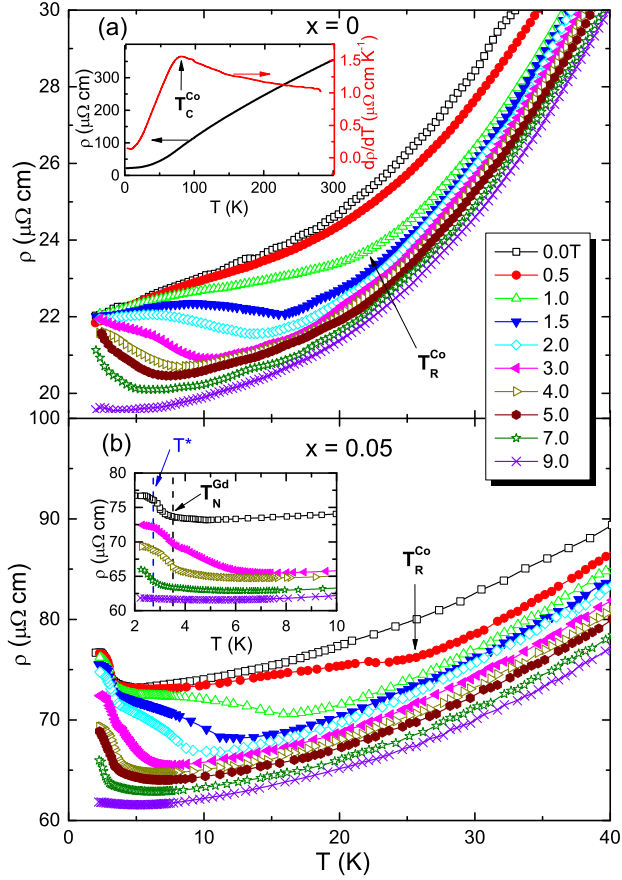}
     \end{center}
     \caption{(Color online) Temperature dependence of the electrical resistivity measured at various magnetic fields for GdCo$_{1-x}$Fe$_x$AsO with $x = 0$ (a) and $0.05$ (b). Inset of (a) plots the zero-field electrical resistivity and its derivative with respect to temperature for $x = 0$ in a temperature range of 2-300 K. Inset of (b) expands the low-$T$ region for $x$ = 0.05. The dashed lines are guides to the eye.}
     \label{fig2}
\end{figure}

Figure 2 plots the electrical resistivity $\rho(T)$ of GdCo$_{1-x}$Fe$_x$AsO ($x$ = 0 and 0.05) at various magnetic fields up to 9 T, showing metallic behavior for both samples. At zero field, we obtained a residual resistivity ratio (RRR) of 16 and 5 for $x$ = 0 and 0.05, respectively. This indicates a good quality of the polycrystalline compounds. In contrast to the GdFeAsO, where a broad maximum attributed to the Fe-AFM order appears around 130 K in the electrical resistivity,\cite{wang2008thorium,tianCe} no distinct resistive anomaly associated with the Co magnetic transitions can be found at zero field in GdCo$_{1-x}$Fe$_x$AsO. Nevertheless, one may see a slope change of the $\rho(T)$ around the Co Curie temperature $T_\textup{C}^\textup{Co}$ of 75 K and 50 K for $x = 0$ and $0.05$, respectively. The $T_\textup{C}^\textup{Co}$ can be tracked from the derivatives of the electrical resistivity with respect to temperature (d$\rho$/d$T$) (see $x = 0$ as an example in the inset of Fig. 2(a)). For $x = 0$, at $\mu_0H \leq 0.5$ T no obvious transition can be found in the electrical resistivity around the temperature where a broad maximum demonstrated in the magnetic susceptibility (see below), which is distinct from the analogue compound, NdCoAsO, where a step-like FM-AFM transition of Co moments is clearly seen around 14 K both in the electrical resistivity and magnetic susceptibility.\cite{mcguire2010magnetic,Ohta2009} However, at $\mu_0 H = 1$ T an obvious transition appears around $T_\textup{R}^\textup{Co} \approx $ 25 K, as indicated by the arrow in Fig. 2(a), which develops into a resistive minimum at higher fields ($\mu_0 H > 1$ T). Such transitions correspond to the spin reorientation of Co moments, being consistent with the maximum in the susceptibility $\chi(T)$ in Fig. 3(a). The $T_\textup{R}^\textup{Co}$ is suppressed to lower temperatures with increasing the magnetic field, and disappears at $\mu_0 H > 7$ T. We note that strong negative magnetoresistance persists down to the temperature we have identified as $T_\textup{R}^\textup{Co}$. Below that temperature, the resistivity returns to the low-temperature limit of the zero-field resistivity. We propose that the FM Co lattice has strong uniaxial anisotropy, with a random orientation of hard $c$-axis in the polycrystalline crystals. In zero field, the grain-to-grain magnetic orientation is random, giving rise to spin-scattering at grain boundaries. In an applied field, the magnetization of neighboring grains is aligned with the magnetic field so long as the Zeeman energy exceeds the anisotropy energy. The difference between low-field and high-field resistance, then, is attributed to spin-dependent grain-boundary scattering. As the temperature decreases, two effects change the balance between Zeeman and anisotropy energy: (1) the anisotropy energy increases with decreasing temperature and (2) the magnetization of Gd exerts an AFM molecular field on Co, reducing the net magnetic field acting on the Co magnetization. Ignoring this latter effect, we note that the grain-boundary resistance is completely suppressed at 9 T. Using the estimate of 0.32 $\mu_\textup{B}$/Co, we place an upper limit on the anisotropy energy to be approximately 3 $\times$ 10$^5$ J/m$^3$, similar to other Co compounds.\cite{Cullity2008} At low temperatures, the Gd-AFM transition is barely visible in the electrical resistivity (see Fig. 2(a)), however, we can track it in the magnetic susceptibility and specific heat (see below). Upon Co/Fe substitution, the Co magnetic transitions decrease to lower temperatures.\cite{tianCe} For $x = 0.05$, the $\rho(T)$ curve shows a distinct transition at $T_\textup{R}^\textup{Co} \approx $ 25 K in a field of 0.5 T, as the arrow shows in Fig. 2(b). Upon increasing the magnetic field, $T_\textup{R}^\textup{Co}$ is also quickly suppressed to lower temperatures before disappearing at $\mu_0 H > 4$ T. However, the $T_\textup{N}^\textup{Gd}$ is robust against the magnetic field, and the resistive upturn at $T_\textup{N}^\textup{Gd}$ is likely associated with a gap opening.\cite{tianCe} We note that the $T_\textup{N}^\textup{Gd}$ of $x = 0.05$ is hardly seen in the electrical resistivity at 9 T, but still can be found in the specific heat. In addition, for $x = 0.05$, a subsequent transition was observed below $T_\textup{N}^\textup{Gd}$ for $\mu_0 H \leq 4$ T (marked as $T^\ast$ in the inset of Fig. 2(b)), which is present in the thermodynamic properties as well (see below). Since this subsequent transition disappears at the field where the $T_\textup{R}^\textup{Co}$ is suppressed, it may arise from a magnetic reorientation of Co moments.

\begin{figure}[tbp]
     \begin{center}
     \includegraphics[width=3.3in,keepaspectratio]{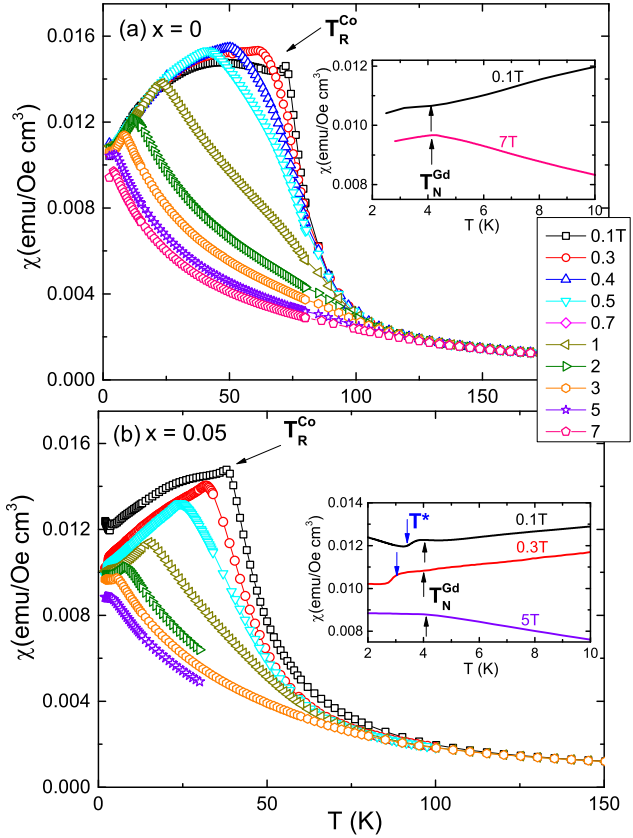}
     \end{center}
     \caption{(Color online) Temperature dependence of the dc magnetic susceptibility $\chi(T)$ at various magnetic fields for GdCo$_{1-x}$Fe$_x$AsO with $x = 0$ (a) and $0.05$ (b). The insets expand the $\chi(T)$ in the low-temperature regions.}
     \label{fig3}
\end{figure}

\begin{figure}[tbp]
     \begin{center}
     \includegraphics[width=3.3in,keepaspectratio]{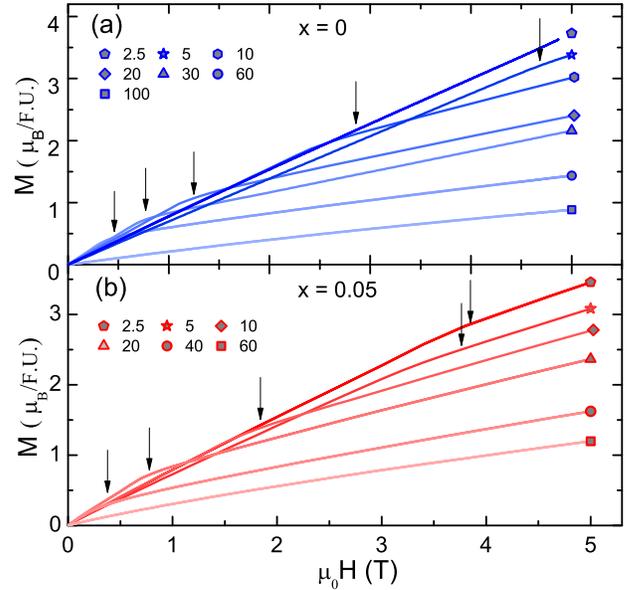}
     \end{center}
     \caption{(Color online) Field dependence of the magnetization $M(H)$ at various representative temperatures for (a) $x$ = 0 and (b) 0.05, respectively. The arrows mark the spin reorientation transitions.}
     \label{fig4}
\end{figure}

The dc magnetic susceptibility for GdCo$_{1-x}$Fe$_x$AsO, measured as a function of temperature at various magnetic fields up to 7 T, is presented in Fig. 3. As an example, here we take GdCoAsO for a detailed analysis (Fig. 3(a)). The magnetic susceptibility increases sharply at a temperature that decreases with $x$, indicating FM nature of this transition. The Curie temperature of Co moments $T_\textup{C}^\textup{Co}$ can be obtained from the derivatives of d$\chi$/d$T$ (see details in Ref. 14). At an extremely small field, the FM transition is sharp in temperature, but it broadens and slightly shifts to higher temperatures with increasing magnetic field.\cite{Chen2012,JunmuSR2011} Similar effects were reported previously in SmCoAsO and NdCoAsO.\cite{Ohta2009} At high temperatures ($T >$ 150 K), the magnetic susceptibility follows a Curie-Weiss law, with the derived effective moment close to the free ion moment of Gd (7.94$\mu_\textup{B}$), indicating the negligible contribution of Co moments.\cite{tianCe} Furthermore, the susceptibility can be modeled within the mean field approach by considering the AFM coupling between the Gd and Co sublattice, giving rise to FIM behavior in the magnetic susceptibility.\cite{tianCe} At temperatures just below $T_\textup{C}^\textup{Co}$, a broad maximum was observed at low fields ($\mu_0 H < 1$ T) that becomes much sharper (peak) at large fields ($\mu_0 H \geq 1$ T), as shown in Fig. 3, being consistent with the reorientation transitions observed in $M(H)$ curves in Fig. 4. As discussed above, such behavior corresponds to the realignment of Co moments away from the field direction due to increasing magnetocrystalline anisotropy at low temperatures. Roughly 1/3 of the magnetic moment should be involved, an estimation that is consistent with the observed decrease in magnetization. For fields above 7 T (5 T for $x$ = 0.05), the local anisotropy is unable to reorient the Co magnetization in hard-axis grains and no sharp decrease can be observed. However, no distinct anomaly can be found in the electrical resistivity and the specific heat for $\mu_0 H < 1$ T, implying that the reorientation process is broad in temperature at low fields. At lower temperatures, the Gd AFM transition ($T_\textup{N}^\textup{Gd}$) can be clearly found in the magnetic susceptibility $\chi(T)$, which is nearly unchanged upon Co/Fe substitution.\cite{tianCe} With increasing the magnetic field, the feature at $T_\textup{C}^\textup{Co}$ is smeared, but  $T_\textup{R}^\textup{Co}$ is rapidly suppressed to lower temperatures and becomes barely visible at $\mu_0 H = 7$ T and 5 T for $x$ = 0 and 0.05, respectively. On the other hand, $T_\textup{N}^\textup{Gd}$ is robust against the magnetic field, as the up arrows indicated in the insets of Figs. 3(a) and (b). For $x$ = 0.05, the subsequent transition below $T_\textup{N}^\textup{Gd}$ also appears in the magnetic susceptibility, as indicated by $T^\ast$ in the inset of Fig. 3(b)(down arrows). Such a subsequent transition is barely visible at $\mu_0 H > 1$ T in the $\chi(T)$, but it can be tracked up to 4 T in $\rho(T)$ (see inset of Fig. 2(b)).

\begin{figure}[tbp]
     \begin{center}
     \includegraphics[width=3.3in,keepaspectratio]{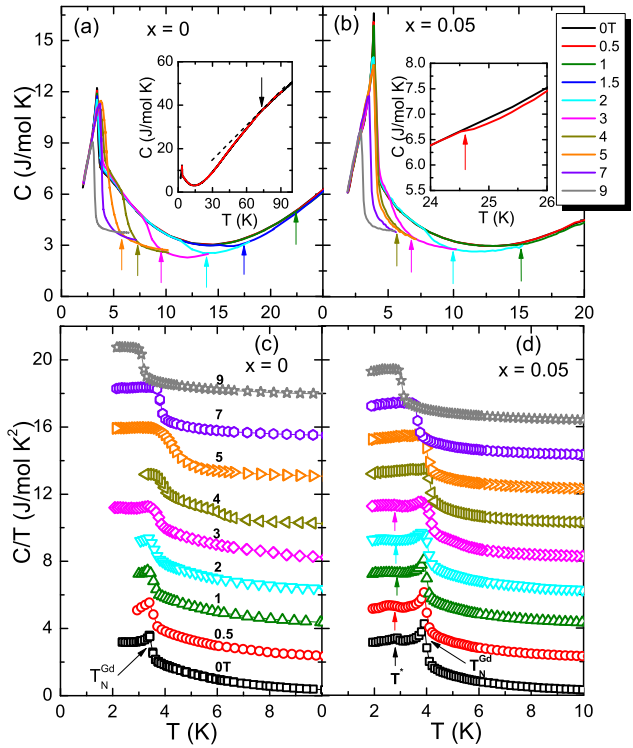}
     \end{center}
     \caption{(Color online) Temperature dependence of the specific heat $C(T)$ at different magnetic fields for GdCo$_{1-x}$Fe$_x$AsO: $x = 0$ (a) and $0.05$ (b). The inset of (a) plots $C(T)$ up to a temperature of 100 K for $x$ = 0 at $\mu_0 H = 0$ T and 0.5 T. The inset of (b) enlarges the temperature range near $T_\textup{R}^\textup{Co}$ for $x$ = 0 at $\mu_0 H = 0$ T and 0.5 T. (c) and (d) plot $C(T)/T$ versus $T$ in the low-T regime for $x$ = 0 and 0.05, respectively. The color of the arrows marks the value of applied magnetic field.}
     \label{fig5}
\end{figure}

We also performed measurements of the field dependent magnetization $M(H)$ up to 5 T at various temperatures (see Fig. 4). In the PM state, the magnetization of both samples is linear in field, e.g., $T $= 100 K and 60 K for $x$ = 0 and 0.05, respectively; while at temperatures below $T_\textup{C}^\textup{Co}$, the $M(H)$ curve changes its slope at a magnetic field that increases with decreasing temperature, indicating a reorientation transition. The applied magnetic field pulls the magnetization of each grain from the local easy direction to the field direction via a first order reorientation transition, i.e., all the moments are polarized to the field direction (see polarized phase in the phase diagram (see below)). Because the orientations of the grains are random, the transition is spread out and becomes continuous. The reorientation transition field $H_\textup{M}$ can be determined from the derivatives of magnetization with respect to magnetic field d$M$/d$H$. The derived $H_\textup{M}$, marked by arrows in Fig. 4, are summarized in the $T-H$ phase diagram (see below). For $x = 0$, the derived $H_\textup{M}$ are consistent with  previous results on GdCoAsO.\cite{Ohta2009,Ohta2012} It is noted that, for $T$ = 2.5 K, the $H_\textup{M}$ of $x = 0$ is larger than 5 T. $H_\textup{M}$ also decreases upon Co/Fe substitution, suggesting that the Fe moments are likely oriented opposite to the Co moments due to internal fields. A scenario involving disorder within the CoAs-layers is also suggested. The Fe atoms acting as impurities in the Co sites, might decrease the Co-Co exchange interaction, making the reorientation transition occur at lower field.

In Fig. 5, we present the temperature dependence of the specific heat for GdCo$_{1-x}$Fe$_x$AsO measured under various magnetic fields up to 9 T. The huge magnetic entropy at $T_\textup{N}^\textup{Gd}$ is attributed to the large spin freedom of Gd-ions ($S\simeq$ Rln(2$s$+1)), while the entropy change associated with Co magnetic transitions is weak, indicating a low-spin configuration of Co-ions. Such behavior agrees with the effective moment derived from the Curie-Weiss analysis of the high-$T$ susceptibility, which is nearly identical to the free-ion moment of Gd, with the contribution of Co moments being negligible. Further experiments, e.g, electron spin resonance (ESR) measurements, would be useful to confirm the spin configurations of Co. In order to demonstrate the Co magnetic transitions, low-$T$ specific heat $C(T)$ is plotted in Figs. 5(a) and (b) for $x$ = 0 and 0.05, respectively. According to the Landau-type theory, the heat capacity would exhibit a jump at the reorientation transition.\cite{Horner1968} For $x = 0$, no obvious anomaly at $T_\textup{R}^\textup{Co}$ can be found in the specific heat for $\mu_0 H < 1$ T, but the slope of $C(T)$ changes at $T_\textup{C}^\textup{Co}$, as shown in the inset of Fig. 5(a). However, when further increasing the magnetic field ($\mu_0 H \geq 1$ T), the $C(T)$ does exhibit a step-like increase at $T_\textup{R}^\textup{Co}$, as the arrows indicate in Fig. 5(a). The reorientation transition temperatures $T_\textup{R}^\textup{Co}$ in the specific heat are highly consistent with those derived from the electrical resistivity $\rho(T)$ and magnetic susceptibility $\chi(T)$ data. The $T_\textup{R}^\textup{Co}$ is also monotonically suppressed on increasing the magnetic field in the $C(T)$, which decreases to 5.7 K at $\mu_0 H = 5$ T before encountering the Gd AFM transition ($T_\textup{N}^\textup{Gd} \approx$ 3.5 K). We note that at $\mu_0 H = 7$ T, it is difficult to determine the $T_\textup{R}^\textup{Co}$ in the specific heat and magnetic susceptibility due to the proximity to $T_\textup{N}^\textup{Gd}$, but it still can be tracked in the electrical resistivity (see Fig. 2(a)). Similar behavior was also observed for $x = 0.05$ except that the $T_\textup{R}^\textup{Co}$ is observable already at $\mu_0 H = 0.5$ T (see inset of Fig. 5(b)), reflecting that Co/Fe substitution decreases the magnetocrystalline anisotropy.

In Figs. 5(c) and (d), we plot the specific heat $C/T$ below 10 K for $x$ = 0 and 0.05 to show the magnetic properties of Gd moments under various magnetic fields. A value of 2 J/mol-K$^2$ is added as an offset. Again, for both compounds, the AFM order of Gd moments is relatively robust against the magnetic field. Similar behavior was also observed in $\rho(T)$ and $\chi(T)$. On the other hand, for $x = 0.05$, a subsequent transition below  $T_\textup{N}^\textup{Gd}$ was also found in the specific heat at $\mu_0H \leq$ 3 T, denoted as $T^\ast$ in Fig. 5(d), being consistent with the $\rho(T)$ and $\chi(T)$ data. This subsequent transition is likely attributed to a further reorientation of Co moments due to the coupling between Co $3d$ and Gd $4f$ electrons.\cite{tianCe}

\begin{figure}[tbp]
     \begin{center}
     \includegraphics[width=3.3in,keepaspectratio]{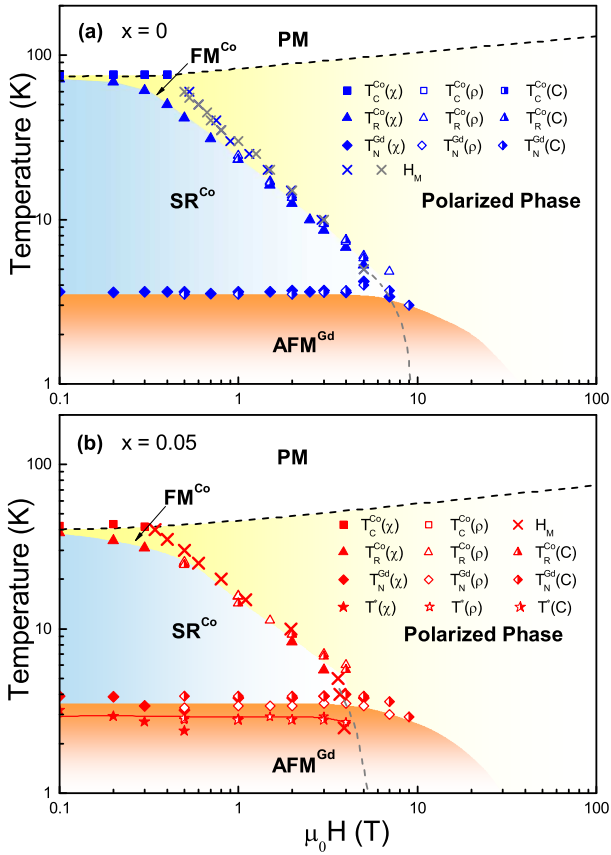}
     \end{center}
     \caption{(Color online) $T$-$H$ phase diagram for GdCo$_{1-x}$Fe$_x$AsO, (a) $x = 0$ and (b) $x = 0.05$. The various symbols denote different types of magnetic transitions determined from the $\rho(T)$, $\chi(T)$ and $C(T)$. The \textcolor[rgb]{0.50,0.50,0.50}{$\bm{\times}$} symbols are taken from Ref. 29. The dashed and solid lines are the guides to the eyes.}
     \label{fig6}
\end{figure}

Figure 6 summarizes all the experimental results in the form of a $T$-$H$ magnetic phase diagram. The application of a magnetic field leads to rich physical properties, which are correspondingly associated with Co $3d$ and Gd $4f$ electrons. At zero field, the Co moments form a long-range FM order with a Curie temperature $T_\textup{C}^\textup{Co} \approx$ 75 K and $T_\textup{C}^\textup{Co} \approx$ 50 K for $x = 0$ and $x = 0.05$, respectively. Just below $T_\textup{C}^\textup{Co}$ and at low field, the AFM coupling between Co and Gd may be sufficient to polarize the Gd moments opposite to the FM Co and the applied field, yielding a narrow FIM regime. As usual, the FM transition gradually broadens as the magnetic field increases (marked by FM$^\textup{Co}$). At temperatures below $T_\textup{C}^\textup{Co}$, when magnetocrystalline anisotropy exceeds the Zeeman energy, it aligns the Co magnetization with the local easy axes of various grains, causing a spin reorientation transition at $T_\textup{R}^\textup{Co}$ (marked by SR$^\textup{Co}$). With increasing magnetic field, the \textquotedblleft gap \textquotedblright between FM$^\textup{Co}$ and SR$^\textup{Co}(H)$ gradually increases before the FM order fades out. At very low temperatures, the Gd moments are ordered antiferromagnetically with a N\'{e}el temperature of $T_\textup{N}^\textup{Gd} \approx 3.5$ K (marked by AFM$^\textup{Gd}$). The $T_\textup{N}^\textup{Gd}$ is nearly unchanged, while the $T_\textup{R}^\textup{Co}$ gradually moves to lower temperature with increasing magnetic field. Since the Gd moments are still ordered antiferromagnetically at 9 T (see Figs. 5(c) and (d)), further study under higher magnetic field is desirable to clarify whether there exists an $f$-electron magnetic critical point in these compounds. It is noted that, for $x = 0.05$, there is another magnetic transition inside the Gd AFM phase, denoted as $T^\ast$ in Fig. 6(b). Such a transition disappears around the field where $T_\textup{R}^\textup{Co}$ is suppressed, so we attribute it to the magnetic reorientation of Co moments. Furthermore, inside the SR$^\textup{Co}$ phase, the applied magnetic field can pull the magnetization of Co grains from the local easy direction to the field direction (marked by Polarized Phase).

In GdCo$_{1-x}$Fe$_x$AsO, we propose that the Co layers exhibit significant magnetocrystalline anisotropy; each grain will have its own local magnetic orientation and there will be spin-flip scattering among the grains. The applied magnetic field can pull the magnetization of each grain from the local easy axes to the field direction via a first-order reorientation transition. As we discussed previously, considering the AFM coupling between the Gd and Co sublattice, the effective field on Co moments can be described as $H_\textup{eff} = H+\lambda M_\textup{Co}-\mu M_\textup{Gd}$ within a mean-field approach.\cite{tianCe} Here the $\lambda$ is the FM Co-Co coupling constant and $\mu$ is the AFM Co-Gd coupling. As we discussed in the resistivity section, at a fixed external field, two factors will modify the reorientation process on decreasing the temperature: (i) the magnetocrystalline anisotropy energy increases and (ii) the effective magnetic field, acting on Co, decreases due to the increased AFM coupling to the Gd paramagnetism ($\mu$). Thus, the magnetic field is unable to suppress the reorientation transition at lower temperatures, causing the Co magnetization to rotate toward the local easy direction in each grain. Consequently, the electrical resistivity approaches the low-field behavior which is dominated by spin-flip scattering at grain boundaries and a step-increase in the specific heat accompanies the reorientation. At low fields, the direction of magnetization changes continuously over a temperature interval of several degrees with no observable changes in the magnitude of magnetization, but it becomes much sharper with increasing magnetic field (see details in Fig. 3). As an example, for $x = 0$, the reorientation transition at low fields ($\mu_0 H < 1$ T) is very broad, the susceptibility exhibits very broad maximum, and no observable transition can be found in the electrical resistivity and specific heat. While for $x = 0.05$, the critical field decreases to 0.5 T, as expected since the Co/Fe substitution in the CoAs-layers weakens the FM Co-Co coupling constant $\lambda$ or increases the magnetocrystalline anisotropy. Therefore, both magnetic field and chemical substitutions can effectively tune the magnetic properties of these compounds, providing the examples for fundamental research and application.

\section{CONCLUSION}

In conclusion, we have synthesized GdCo$_{1-x}$Fe$_x$AsO ($x = 0$ and 0.05) polycrystalline samples and systematically investigated the field-tuning effects on the electrical resistivity, magnetization and specific heat. A rich $T$-$H$ phase diagram has been constructed. It is found that the magnetic field can effectively tune the magnetic properties of Co $3d$ electrons, while the Gd $4f$ electrons are insensitive to the applied magnetic field. Further investigations on single crystals are badly needed in order to clarify the role of magnetocrystalline anisotropy. To achieve that, one still needs to overcome the difficulties of growing sizable single crystals. Moreover, it is also interesting to extend the measurements to higher magnetic fields and lower temperatures to study the possible magnetic critical behaviors.

\begin{acknowledgments}
We acknowledge the fruitful discussions with Marcelo Jaime. Work at Zhejiang University is supported by the National Basic Research Program of China (2011CBA00103),the National Science Foundation of China (No. 11174245) and the Fundamental Research Funds for the Central Universities. Work at Los Alamos National Lab was performed under the auspices of the US DOE. Work at The University of Texas at Dallas is supported by the AFOSR grant (No.FA9550-09-1-0384) on search for novel superconductors.
\end{acknowledgments}

\end{document}